\documentclass[twocolumn,showpacs,preprintnumbers]{revtex4}

\usepackage{graphicx}% Include figure files
\usepackage{dcolumn}% Align table columns on decimal point
\usepackage{bm}% bold math
\begin{document}

%\draft

\title{Pressure Dependence of Fragile-to-Strong Transition and a Possible
Second Critical Point in Supercooled Confined Water}

\author{Li Liu,$^{1}$ Sow-Hsin Chen,$^{1}$\footnote{Author to whom correspondence should be
addressed. Electronic mail: sowhsin@mit.edu} Antonio
Faraone,$^{1}$ Chun-Wan Yen,$^{2}$ and Chung-Yuan Mou$^{2}$ }
\address{$^{1}$Department of Nuclear Science and Engineering, Massachusetts Institute
of Technology, Cambridge, Massachusetts 02139 USA\\
$^{2}$Department of Chemistry, National Taiwan University, Taipei
106, Taiwan}

\date{\today}

\begin{abstract}
By confining water in nano-pores of silica glass, we can bypass
the crystallization and study the pressure effect on the dynamical
behavior in deeply supercooled state using neutron scattering. We
observe a clear evidence of a cusp-like fragile-to-strong (F-S) dynamic
transition. Here we show that the transition temperature decreases
steadily with an increasing pressure, until it intersects the
homogenous nucleation temperature line of bulk water at a pressure
of 1600 bar. Above this pressure, it is no longer possible to
discern the characteristic feature of the F-S
transition. Identification of this end point with the possible
second critical point is discussed.
\end{abstract}


\maketitle

Water is a continuing source of fascination to scientists because
of its many counterintuitive low-temperature properties. Although
the stable form of water at sufficiently low temperature is
inevitably crystalline, liquid water can also exist in a
metastable supercooled state far below the melting point. One of
the most intriguing questions related to the unusual properties of
supercooled water is whether two critical points may exist in a
single component liquid~\cite{Deben1}.

It has been known since 1970s that a number of thermodynamic
response functions of supercooled water, notably the isothermal
compressibility $K_T$ and the constant-pressure specific heat
$C_P$, show a power-law divergence behavior at a singular
temperature, experimentally determined to be 228~K at the ambient
pressure~\cite{Speedy,Angell1}. Concurrently, the transport
properties, such as the shear viscosity $\eta$ and the inverse self-diffusion
constant D, diverge according to power-laws
toward the same singular temperature~\cite{Pri,Osi}. The anomalies
of the thermodynamic quantities become plausible if one postulates
the existence of a second low-temperature critical point at about
228~K and at somewhat elevated pressure~\cite{Poole}. On the other
hand, the transport coefficient anomalies are reminiscent of the
dynamical behavior of a supercooled liquid near the so-called
kinetic glass transition temperature, predicted by Mode-Coupling
theory (MCT)~\cite{Gotze}.

Search for the predicted~\cite{Poole} first-order liquid-liquid
transition line and its end point, the second low-temperature
critical point~\cite{Deben1,Deben2} in water, has been hampered by
intervention of the homogenous nucleation process, which takes
place at 235~K at the ambient pressure. However, by confining
water in nano-pores of mesoporous silica MCM-41-S with cylindrical
pores of 14~\AA~diameter, we have been able to study the dynamical
behavior of water in a temperature range down to 160~K, without
crystallization. Using high-resolution Quasi-Elastic Neutron
Scattering (QENS) method and Relaxing-Cage Model (RCM) for
the analysis, we determine the temperature and pressure dependences of
the average translational relaxation time $\left\langle \tau_T
\right\rangle$ for the confined supercooled water.

Micellar templated mesoporous silica matrices MCM-41-S, which have
1-D cylindrical pores arranged in 2-D hexagonal arrays, were
synthesized by following a similar method for synthesizing
MCM-48-S previously~\cite{PCShih}. Using the short chain cationic
surfactants, C$_{12}$TMAB, as templates and adding $\beta$-type
zeolite seeds as silica source~\cite{YLiu}, we obtain MCM-41-S (S
denotes seed) with smaller pore sizes and stronger silica walls
than the traditional MCM-41. In this pressure
experiment, we chose the silica with a pore diameter of
14~\AA~because for the fully hydrated sample the differential
scanning calorimetry (DSC) data shows no freezing peak down to
160~K. The sample is then hydrated by exposing to water vapor in a
closed chamber until it reaches the full hydration level of 0.5
gram $H_2O/1$ gram Silica.

High-resolution QENS spectroscopy method is used to determine the
temperature and pressure dependences of $\left\langle \tau_T
\right\rangle$ for the confined water. Because neutrons can easily
penetrate the thick-wall high-pressure cell and because it is
predominantly scattered by hydrogen atoms in water, rather than by
the matrices containing it, incoherent QENS is an appropriate tool
for our study. Using two separate high-resolution QENS
spectrometers, we are able to measure the translational
-relaxation time from 0.2~$ps$ to 10,000~$ps$ over the temperature
and pressure range. The high-pressure experiments were performed
at both the High-Flux Backscattering (HFBS) and the Disc-Chopper
Time-of-Flight (DCS) spectrometers in the NIST Center for Neutron
Research (NIST NCNR). The two
spectrometers used to measure the spectra have two widely
different dynamic ranges (for the chosen experimental setup), one
with an energy resolution of 0.8~$\mu$eV (HFBS) and a dynamic
range of $\pm$11~$\mu$eV~\cite{Meyer}, and the other with an energy
resolution of 20~$\mu$eV (DCS) and a dynamic range of $\pm$0.5~meV
~\cite{JRC} in order to be able to extract the broad range of
relaxation times from the measured spectra. The same high pressure
system, including specially designed aluminium pressure cell, was
used on the two instruments. Helium gas, the pressure-supplying
medium, fills the whole sample cell, and applies pressure to the
fully hydrated sample. The experiment at each pressure was done
with a series of temperatures, covering both below and above the
transition temperature. Altogether, 1000 spectra were collected,
spanning 9 pressures: ambient, 100, 200, 400, 800, 1200, 1600,
2000, and 2400 bars.

QENS experiments measure the Fourier transform of the Intermediate
Scattering Function (ISF) of the hydrogen atoms, $F_H(Q,t)$, of
water molecules in the pores of the silica matrix. Molecular
Dynamics (MD) simulations have shown that the ISF of both
bulk~\cite{Gallo1} and confined~\cite{Gallo2} supercooled water
can be accurately described as a two-step relaxation: a short-time
Gaussian-like (in-cage vibrational) relaxation followed by a
plateau and then a long-time (time $>$ 1.0~$ps$) stretched
exponential relaxation of the cage. The so-called Relaxing Cage
Model (RCM)~\cite{ChenPRE}, which we use for data analysis, models
closely this two-step relaxation and has been tested extensively
against bulk and confined supercooled water through MD and
experimental data~\cite{Gallo1,Gallo2,ChenPRE}. By considering
only the spectra with wave vector transfer $Q < 1.1$~\AA$^{-1}$,
we can safely neglect the contribution from the rotational motion
of water molecule in ISF~\cite{ChenPRE}. The RCM describes the
translational dynamics of water at supercooled temperature in
terms of the product of two functions:

\begin{eqnarray}
F_H\left(Q,t\right)&\approx& F_T\left(Q,t\right)
=F^S\left(Q,t\right)exp\left[-\left(t/\tau_T(Q)\right)^\beta\right], \nonumber \\
\tau_T\left(Q\right)&=&\tau_0\left(0.5Q\right)^{-\gamma},
\left\langle\tau_T\right\rangle=\tau_0\Gamma\left(1/\beta\right)/\beta,
\label{decoupling}
\end{eqnarray}
where the first factor, $F^S\left(Q,t\right)$, represents the
short-time vibrational dynamics of the water molecule in the cage.
This function is fairly insensitive to temperature variation, and
thus can be calculated from MD simulation. The second factor, the
$\alpha$-relaxation term, contains the stretch exponent $\beta$,
and the $Q$-dependent translational relaxation time
$\tau_T\left(Q\right)$, which is a strong function of temperature.
The latter quantity is further specified by two phenomenological
parameters $\tau_0$ and $\gamma$, the exponent controlling the
power-law $Q$-dependence of $\tau_T\left(Q\right)$.
$\left\langle\tau_T\right\rangle$ is a $Q$-independent
quantity where $\Gamma$ is the gamma function. It essentially gives a measure of the
structural relaxation time of the hydrogen-bond cage surrounding a
typical water molecule. The temperature dependence
of the translational relaxation time at each pressure is then
calculated from three fitted parameters, $\tau_0$, $\beta$, and
$\gamma$, by analyzing a group of quasi-elastic peaks at different
$Q$ values simultaneously. For this analysis, we chose seven spectra
for data taken at HFBS and eleven spectra for data taken at DCS,
at each temperature.

\begin{figure}[tbp]
\begin{center}
\includegraphics[width=8.5 cm]{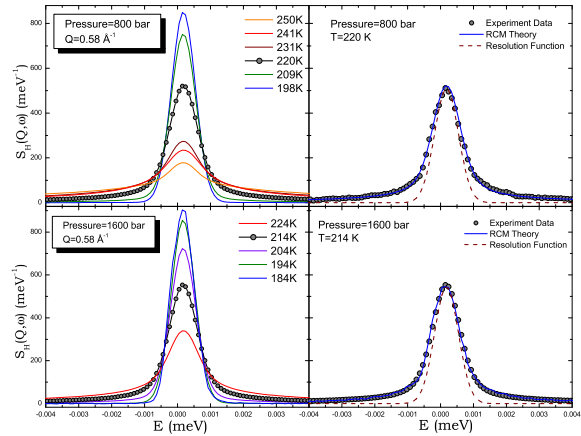}
\end{center}
\caption{{{{{{{{{{{{\protect\small Figs.\ref{fig1}A and
\ref{fig1}C (left panels) show QENS spectra measured at $Q=0.58$~
\AA$^{-1}$, at two pressures, 800 bar and 1600 bar, and at a
series of temperatures. Figs.\ref{fig1}B and \ref{fig1}D (right
panels) show the RCM analysis of one of the spectrum from each
pressure. The resolution function in each case is shown
by a dashed line. %
}}}}}}}}}}}} \label{fig1}
\end{figure}

\begin{figure}[tbp]
\begin{center}
\includegraphics[width=7.5 cm]{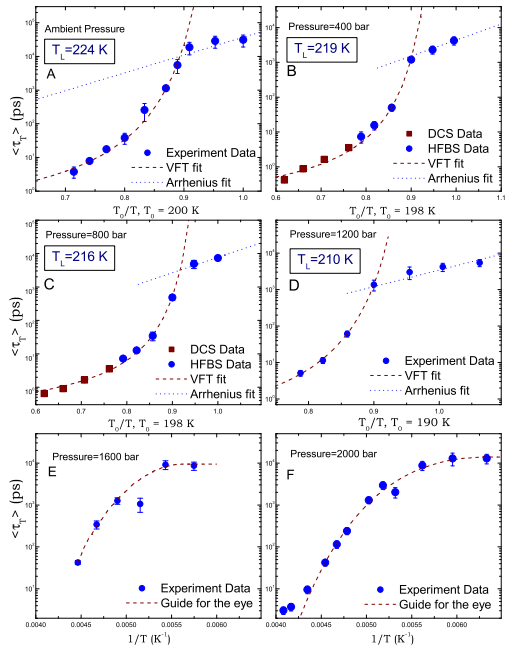}
\end{center}
\caption{{{{{{{{{{{{\protect\small Temperature dependence of
$\langle \tau_T \rangle$ plotted in $log(\langle \tau_T \rangle)$
vs $T_0/T$ or $1/T$. Data from ambient pressure, 400, 800, 1200, 1600,
and 2000 bars are shown in panels A, B, C, D, E, and F,
respectively. %
}}}}}}}}}}}} \label{fig2}
\end{figure}

\begin{figure}[tbp]
\begin{center}
\includegraphics[width=7.5 cm]{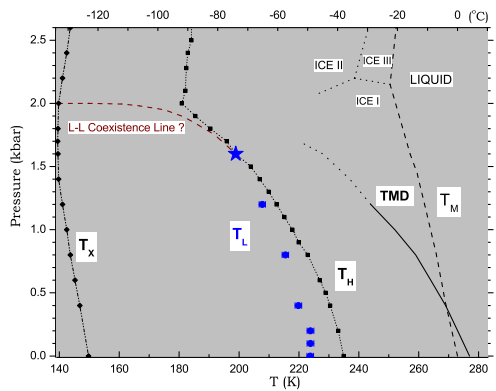}
\end{center}
\caption{{{{{{{{{{{{\protect\small The pressure
dependence of the F-S dynamic transition
temperature, $T_L$, plotted in the P-T plane (solid circles). Also
shown are the homogeneous nucleation temperature line, denoted as
$T_H$~\cite{Kanno}, crystallization temperatures of amorphous
solid water~\cite{Stanley00}, denoted as $T_X$, and the
temperature of maximum density line, denoted as
TMD~\cite{Angell3}.%
}}}}}}}}}}}} \label{fig3}
\end{figure}

We show in Fig.\ref{fig1}, as an example, two complete sets
(temperature series) of QENS area-normalized spectra. The
broadening of the quasi-elastic peaks becomes more and more
noticeable as temperature increases. In Fig.\ref{fig1}A, we may
notice, from shoulders of these spectral lines, that two groups of
curves, 231-250~K and 198-209~K, are separated by the curve at a
temperature of 220~K. This visual information reinforces the
result of the detail analysis shown in Fig.\ref{fig2}, that there
is an abrupt dynamical transition at $T_L = 216$~K.
Fig.\ref{fig1}B shows the RCM analysis of the spectrum taken at $T
= 220$~K, close to the transition temperature. On the other hand
in Fig.\ref{fig1}C, the spectra at pressure 1600 bar show a rather
smooth variation with temperature, indicating that there is no
sharp transition. This pressure corresponds to the end point of
the line of F-S transitions shown in Fig.\ref{fig3}.
Fig.\ref{fig1}D is, again, an RCM analysis of the spectrum taken
at $T =214$~K of this pressure. RCM, as one can see, reproduces
the experimental spectral line shapes of confined water well. The
broadening of the experimental data over the resolution function,
shown in Figs.\ref{fig1}B and Fig.\ref{fig1}D, leaves enough
dynamic information to be extracted by RCM.

The behavior of shear viscosity $\eta$ or equivalently the
structural relaxation time $\tau$ of a supercooled liquid
approaching its glass transition temperature is called `fragile'
when it varies according to the so-called Vogel-Fulcher-Tammann
(VFT) law; and the behavior is called `strong' when $\eta$ or
$\tau$ obeys Arrhenius law~\cite{Angell2}. For water, a fragile
liquid at room temperature and at moderately supercooled
temperatures, Ito and coworkers~\cite{Ito} proposed that a
`fragile-to-strong' transition would occur at around 228~K, based
on a thermodynamic argument. The F-S transition in a molecular
liquid like water may be interpreted as a variant of kinetic glass
transition predicted by the ideal MCT~\cite{Gotze}, where the real
structural arrest transition is avoided by an activated hoping
mechanism below the transition.

In Fig.\ref{fig2}, we report the temperature variation of
$\left\langle\tau_T\right\rangle$ for water molecules as a
function of pressure. It is seen that Figs. \ref{fig2}A,
\ref{fig2}B, \ref{fig2}C and \ref{fig2}D show clearly a transition
from a VFT law: $\left\langle\tau_T\right\rangle=\tau_1
exp\left[DT_0/(T - T_0)\right]$, where $D$ is a constant providing
the measure of fragility and $T_0$, the ideal glass transition
temperature, to an Arrhenius law:
$\left\langle\tau_T\right\rangle=\tau_1 exp\left[E_A/RT\right]$,
where $E_A$ is the activation energy for the relaxation process
and $R$, the gas constant. This transition of VFT to Arrhenius
behavior, previously observed at ambient pressure~\cite{Faraone},
is the signature of a F-S dynamic transition predicted by Ito $et$
$al$~\cite{Ito}. In this paper, we show the extension of this
transition into finite pressures. The transition temperature,
$T_L$, as the crossing point of the VFT law and Arrhenius law, is
calculated by $1/T_L = 1/T_0 - (Dk_B)/E_A$. However, in
Figs.\ref{fig2}E and \ref{fig2}F, the cusp-like transition becomes
rounded off and there is no clear-cut way of defining the F-S
transition temperature. Note that in Fig.\ref{fig2}F, there is
still a hint of fragile behavior at high enough temperature.

Summarizing all the results, we show in a P-T plane, in
Fig.\ref{fig3}, the observed pressure dependence of $T_L$ and its
estimated continuation, denoted by a dash line, in the pressure
region where no clear-cut F-S transition is observed. One should
note that the $T_L$ line has a negative slope, parallel to TMD
line, indicating a lower density liquid on the lower temperature
side. This $T_L$ line also approximately tracks the $T_H$ line,
and terminates in the upper end when intersecting the $T_H$ line
at 1600 bar and 200~K, at which point the character of the dynamic
transition changes. We shall discuss the significance of this
point later on. A special feature of the $T_L$ line at the lower
end should be noted as well. The line essentially becomes vertical
after around 200 bar and the transition temperature approaches a
constant value of $\sim$225~K.

Since $T_L$ determined experimentally is a dynamic transition
temperature, it is natural to question whether the system is in a
liquid state on both sides of the $T_L$, and if so, what would the
nature of the high-temperature and low-temperature liquids be?
Sastry and Angell have recently shown by a MD simulation that at a
temperature $T\approx 1060$~K (at zero pressure), below the
freezing point 1685~K, the supercooled liquid silicon undergoes a
first-order liquid-liquid phase transition, from a fragile, dense
liquid to a strong, low-density liquid with nearly tetrahedral
local coordination~\cite{Sastry}. Prompted by this finding, we may
like to relate, in some way, our observed $T_L$ line to the
liquid-liquid transition line, predicted by MD simulations of
water~\cite{Poole} and speculating on the possible location of the
low-temperature critical point.

According to our separate inelastic neutron scattering
experiments, which measure the librational density of states of
water contained in 18~\AA~pore size MCM-41-S, water remains in
disordered liquid state both above and below the F-S
transition at ambient pressure. Furthermore, our analysis of the
F-S transition for the case of ambient pressure
indicates that the activation energy barrier for initiating the
local structural relaxation is $E_A = 4.89$~Kcal/mol for the
low-temperature strong liquid. Yet, previous inelastic scattering
experiments of stretch vibrational band of water~\cite{Ricci}
indicate that the effective activation energy of breaking a
hydrogen bond at 258~K (high-temperature fragile liquid) is
3.2~Kcal/mol. Therefore, it is reasonable to conclude that the
high-temperature liquid corresponds to the high-density liquid
(HDL) where the locally tetrahedrally coordinated hydrogen bond
network is not fully developed, while the low-temperature liquid
corresponds to the low-density liquid (LDL) where the more open,
locally ice-like hydrogen bond network is fully
developed~\cite{Soper}.

It is appropriate now to address the possible location of the
second critical point~\cite{Poole}. Above the critical temperature
$T_C$ and below the critical pressure $p_C$, we expect to find a
one-phase liquid with a density $\rho$, which is constrained to
satisfy an equation of state: $\rho = f (p, T)$. If an experiment
is done by varying temperature $T$ at a constant pressure $p < p_C$, $\rho$
will change from a high-density value (corresponding to HDL) at
sufficiently high temperature to a low-density value
(corresponding to LDL) at sufficiently low temperature. Since the
fragile behavior is associated with HDL and the strong behavior
with LDL, we should expect to see a clear F-S
transition as we lower the temperature at this constant $p$.
Therefore, the cusp-like F-S transition we observed
should then occur when we cross the so-called Widom line in the
one-phase region~\cite{Private01}. On the other hand, if the
experiment is performed in a pressure range $p > p_C$,
corresponding to the two-phase region and crossing the Liquid-Liquid (L-L)
coexistence line, the system will be consisting of mixture of
different proportions of HDL and LDL as one varies T. In this
latter case, $\langle \tau_T \rangle$ vs. $1/T$ plot will not show
a clear-cut F-S transition (the transition will be
washed out) because the system is in a mixed state. The above
picture would then explain the dynamical behavior we showed in
Fig.\ref{fig2}.  In Fig.\ref{fig2}, a clear F-S
transition is observed up to 1200 bar and beyond 1600 bar the
transition is rounded off. From this observation, the reasonable
location of the L-L critical point is estimated to be at
$p_C = 1600 \pm 400$~bar and $T_C = 200 \pm 10$~K, shown by a star
in Fig.\ref{fig3}.

Additionally, in a recent MD simulation using Jagla Model
Potential by Xu $et$ $al$~\cite{Private02}, a small peak was found
in the specific heat $C_P$ when crossing the Widom line at a
constant $p$. Meanwhile, Maruyama $et$ $al$ conducted an
experiment on adiabatic calorimetry of water confined within
nano-pores of silica gel~\cite{Maruyama}. It was found that water
within 30~\AA~pores was well prevented from crystallization, and
also showed a small $C_P$ peak at 227~K at ambient pressure. This
experimental result further supports that the F-S transition we
observed at 225~K at ambient pressure is caused by the crossing of
the Widom line in the one-phase region above the critical
point~\cite{Private02}.

\begin{center}
{\bf Acknowledgement }
\end{center}

The research at MIT is supported by DOE Grants DE-FG02-90ER45429
and 2113-MIT-DOE-591. This work utilized facilities supported in
part by the National Science Foundation under Agreement No.
DMR-0086210. This work was also supported by the National Science
Council of Taiwan. Technical supports in measurements from E.
Mamontov, J. Leao, and J.R.D. Copley at NIST NCNR are greatly
appreciated. We acknowledge an enlightening discussion with C.A.
Angell on this subject. We benefited from affiliation with
EU-Mari-Curie Research and Training Network on Arrested Matter.


\begin{references}
\bibitem{Private02} Private communication with H.E. Stanley.
The MD simulation results using Jagla Model Potential will be
published in a paper entitled: Connection between the Widom line and the strong-to-fragile transition.
\end{references}
\end{document}